\documentstyle[12pt]{article}
\makeatletter

\if@twoside
\else
\fi
\ifcase \@ptsize
\or % mods for 11 pt
\or % mods for 12 pt
\fi

\def\section{\@startsection {section}{1}{\z@}{-3.5ex plus -1ex minus
 -.2ex}{2.3ex plus .2ex}{\bf\raggedright}}
\def\subsection{\@startsection{subsection}{2}{\z@}{-3.25ex plus -1ex minus
 -.2ex}{1.5ex plus .2ex}{\sc\raggedright}}
\def\bibliography#1{\if@filesw\immediate\write\@auxout
  {\string\bibstyle{npb}}\fi
  \if@filesw\immediate\write\@auxout{\string\bibdata{#1}}\fi
  \@input{\jobname.bbl}}
\def\thebibliography#1{\section{References\@mkboth
 {REFERENCES}{REFERENCES}}\list
 {[\arabic{enumi}]}{\settowidth\labelwidth{[#1]}\leftmargin\labelwidth
 \advance\leftmargin\labelsep \itemsep=0pt
 \usecounter{enumi}}
 \def\newblock{\hskip .11em plus .33em minus .07em}
 \sloppy\clubpenalty4000\widowpenalty4000
 \sfcode`\.=1000\relax}
\def\@citex[#1]#2{%
\if@filesw \immediate \write \@auxout {\string \citation {#2}}\fi
\@tempcntb\m@ne \let\@h@ld\relax \def\@citea{}%
\@cite{%
  \@for \@citeb:=#2\do {%
    \@ifundefined {b@\@citeb}%
      {\@h@ld\@citea\@tempcntb\m@ne{\bf ?}%
      \@warning {Citation `\@citeb ' on page \thepage \space undefined}}%
      {\@tempcnta\@tempcntb \advance\@tempcnta\@ne%
      \@tempcntb\number\csname b@\@citeb \endcsname \relax%
      \ifnum\@tempcnta=\@tempcntb %   Number follows previous--hold on to it
        \ifx\@h@ld\relax%
          \edef \@h@ld{\@citea\csname b@\@citeb\endcsname}%
        \else%
          \edef\@h@ld{\ifmmode{-}\else--\fi\csname b@\@citeb\endcsname}%
        \fi%
      \else%   %  non-successor--dump what's held and do this one
        \@h@ld\@citea\csname b@\@citeb \endcsname%
        \let\@h@ld\relax%
      \fi}%
    \def\@citea{,\penalty\@highpenalty\,}%
  }\@h@ld%
}{#1}}
\newif\ifabstract \abstractfalse
\newif\ifconsolidatetitle \consolidatetitlefalse
\def\ds@consolidatetitle{\consolidatetitletrue}
\gdef\@publabel{\hfil}
\gdef\@pubdate{\null}
\gdef\@pubnumber{\null}
\gdef\@author{\null}
\gdef\@title{\null}
\gdef\@abstract{\null}
\long\def\pubdate#1{\gdef\@pubdate{#1}}
\long\def\pubnumber#1{\gdef\@pubnumber{#1}}
\long\def\publabel#1{\gdef\@publabel{#1}}
\long\def\author#1{\gdef\@author{#1}}
\long\def\title#1{\gdef\@title{#1}}
\long\def\abstract#1{\abstracttrue\gdef\@abstract{#1}}
\def\titlerelax{
\let\maketitle\relax
\let\settitleparameters\relax
\let\consolidatetitle\relax
\let\inittitlepage\relax
\let\finishtitlepage\relax
\let\titlepagecontents\relax
\let\multithanks\relax
\let\titlebaselines\relax
\let\@makepub\relax
\let\@maketitle\relax
\let\@makeauthor\relax
\let\@makeabstract\relax
\let\@maketitlenote\relax
\let\thanks\relax
\let\titlerelax\relax}
\def\titleclean
{\gdef\@titlenote{}
\gdef\@abstract{}
\gdef\@author{}
\gdef\@title{}
\gdef\@pubdate{}\gdef\@pubnumber{}\gdef\@publabel{}
\gdef\@dpublabel{}
}
\def\@makepub{\vbox to \z@{\hbox to \textwidth{\hfill
\@publabel \hfill
\llap{\parbox[t]{0.25\textwidth}{\raggedleft\@pubnumber}}}%
\vss}}
\def\@maketitle{\vskip 60pt \begin{center}
 {\LARGE \@title \par}
 \end{center}}
\def\@makeauthor{{%
\def\and{\smallskip {\normalsize \rm and\smallskip }}
\def\And{\medskip {\normalsize \rm and\\}\medskip}
\long\def\address##1{{\def\and{\\and\\}\medskip
				{\small \it \\##1\\}
}}
{\centering
 \vskip 3em
 \large \lineskip .75em
 \@author}
 \par}}
\def\@makedate{\vskip 1.5em
 {\raggedright \small \noindent\@pubdate \par}}
\def\@makeabstract{\vskip 1.5em
{\small
\begin{center}
{\bf ABSTRACT\vspace{-.5em}\vspace{0pt}}
\end{center}
\quotation \@abstract \endquotation}}
\def\@consolidatetitle{{
\thispagestyle{empty}
\@makepub
\setlength{\parskip}{0pt}
\null
\vskip 7mm
\nointerlineskip
\@maketitle
\vskip 5ex
\@makeauthor
\vskip 4 ex
\@makeabstract
\vskip 5 ex
}}

\def\maketitle{\ifconsolidatetitle\@consolidatetitle
\else
	\titlepage
	\let\footnotesize\small \setcounter{page}{0}
	\@makepub
	\vfil
	\@maketitle
	\@makeauthor
	\vfil
	\ifabstract\@makeabstract\fi
	\@thanks
	\vfil
	\@makedate
	\if@restonecol\twocolumn\fi \eject
\fi
	\titlerelax \titleclean
	\setcounter{footnote}{0}
}
\def\cF{{\cal F}}

\def\cH{{\cal H}}

\let\a=\alpha
\let\b=\beta

\let\l=\lambda

\let\z=\zeta
\def\bra#1.{{\langle#1|}}
\def\cdd{{\cdot}}

\def\cont{\nonumber\\*&&\mbox{}}
\def\en{\end{equation}}
\def\enn{\end{eqnarray}}
\def\eq{\begin{equation}}
\def\eqq{\begin{eqnarray}}

\def\ket#1.{{|#1\rangle}}

\def\mno{{\textstyle {\circ\atop\circ}}}

\def\rank{\mathop{\rm rank}\nolimits}
\def\rank{\rm rank\,}
\def\Tr{\mathop{\rm Tr}\nolimits}
\def\vac{\vec 0}
\def\vec#1{\vert #1 \rangle}

\def\WA{\mathop{\it WA}\nolimits}
\def\wan{$\WA_n$ }

\def\WB{\mathop{\it WB}\nolimits}

\def\WD{\mathop{\it WD}\nolimits}

\def\eqq{\begin{eqnarray}}
\def\enn{\end{eqnarray}}

\ifcase\@ptsize
 \font\tenmsa=msam10
 \font\sevenmsa=msam7
 \font\fivemsa=msam5
 \font\tenmsb=msbm10
 \font\sevenmsb=msbm7
 \font\fivemsb=msbm5
\or
 \font\tenmsa=msam10 scaled \magstephalf
 \font\sevenmsa=msam8
 \font\fivemsa=msam6
 \font\tenmsb=msbm10 scaled \magstephalf
 \font\sevenmsb=msbm8
 \font\fivemsb=msbm6
\or
 \font\tenmsa=msam10 scaled \magstep1
 \font\sevenmsa=msam8
 \font\fivemsa=msam6
 \font\tenmsb=msbm10 scaled \magstep1
 \font\sevenmsb=msbm8
 \font\fivemsb=msbm6
\fi

\newfam\msafam
\newfam\msbfam
\textfont\msafam=\tenmsa  \scriptfont\msafam=\sevenmsa
  \scriptscriptfont\msafam=\fivemsa
\textfont\msbfam=\tenmsb  \scriptfont\msbfam=\sevenmsb
  \scriptscriptfont\msbfam=\fivemsb

\def\hexnumber@#1{\ifnum#1<10 \number#1\else
 \ifnum#1=10 A\else\ifnum#1=11 B\else\ifnum#1=12 C\else
 \ifnum#1=13 D\else\ifnum#1=14 E\else\ifnum#1=15 F\fi\fi\fi\fi\fi\fi\fi}

\def\msa@{\hexnumber@\msafam}
\def\msb@{\hexnumber@\msbfam}
\mathchardef\therefore="3\msa@29
\mathchardef\Subset="3\msa@62
\mathchardef\Supset="3\msa@63
\mathchardef\Cup="2\msa@64

\mathchardef\Cap="2\msa@65

\def\checkmark{\mathhexbox\msa@58 }
\mathchardef\hbar="0\msb@7E
\mathchardef\backepsilon="3\msb@7F
\def\Bbb{\ifmmode\let\next\Bbb@\else
 \def\next{\errmessage{Use \string\Bbb\space only in math mode}}\fi\next}
\def\Bbb@#1{{\Bbb@@{#1}}}
\def\Bbb@@#1{\fam\msbfam#1}

\newcount{\q}
\thicklines
\def\dynkina#1{
\q=#1 \advance\q by -1
\put(00,0){\circle{2}}
\multiput(01,0)(10,0){\q}{\line(1,0){8}}
\multiput(10,0)(10,0){\q}{\circle{2}}
}
\makeatother

\begin{document}

\pubnumber{DUR-CPT 92-15}
\pubdate{Mar 1992}
\title{A note on W--algebra Realisations}
\author{G. M. T. WATTS\thanks{Email: {\tt G.M.T.Watts@UK.AC.DURHAM}}\address{
Department of Mathematical Sciences, \\
University of Durham, South Road,
Durham, DH1 3LE, U.K.}
}

\abstract{
We provide a general description of realisations of W--algebras in
terms of smaller W--algebras and free fields. This is based on the
definition of the W--algebra as the commutant of a set of  screening
charges. This is conjectured to be related to partial gauge-fixings in the
Hamiltonian reduction model.}

\maketitle

\section{Introduction}

In this note we hope to clarify some recent results on W--algebra
realisations. W--algebras are non-linear generalisations of the
Virasoro algebra which arise as symmetry algebras in two-dimensional
conformal field theory. The first such extension was constructed
abstractly by Zamolodchikov \cite{Zamo1}, and it became clear that
such algebras could be associated to any semi-simple simply-laced Lie
algebra, and
constructions could be found in free fields
\cite{FLuk1,FLuk2}, in GKO coset models \cite{BBSS}. and in
%terms of
the fields in $\hat g$ where $g$ is the maximally non-compact
real form of the algebra \cite{BFFOW}. However, recently
constructions have been found of W--algebras in terms of
W--algebras for groups of lower rank, and free fields
\cite{Roma1,LPS}. These are based heavily on the free
field construction of \cite{FLuk2}. In this note we
%show that an
analyse
this free field construction further enabling us to give a
comprehensive set of realisations of any particular W--algebra in
terms of lower rank W--algebras and free fields.
Firstly we review the free field construction, and then the known
realisations in terms of the Virasoro algebra and free fields.  We
then analyse these models using the free field construction an
d screening charges. We end with a short example and some discussion
on extensions to the affine case and the connection with Hamiltonian
reduction.

\section{Free field construction of W--algebras}

A Quantum W--algebra can be defined in terms of a number of basic
fields $\{W^a\}$ and their operator product expansions \cite{Watt3}.
One of these
fields, denoted by $L(a)$, forms a closed operator product algebra
with itself, corresponding to the Virasoro algebra
The value of the central charge $c$ of this Virasoro algebra is
usually the only free parameter in the W--algebras, the other
structure constants being functions of $c$. The other fields
are quasi-primary fields with respect to this Virasoro algebra of
given conformal weight.
Some progress has been made by considering the consistency requirement
in conjunction with the abstract definition to
show that this imposes strict limits on the conformal spins of the
other basic fields. In particular W--algebras which have a few basic
fields of low conformal weight have received particular attention and
it has proven possible to show that the W--algebras with fields of
weights
$\{2,3\}, \{2,3,4\}, \{2,4\}, \{2,6\}$ are consistent and unique and
conversely that algebras with fields of spins $\{2,5\}, \{2,7\}$
cannot be consistently defined for arbitrary $c$
\cite{Bouw3,BFKNRV1,KWat1}. However most progress has been made by
considering a particular realisation in terms of free massless fields.
This was the form in which the original W--algebra was first
constructed \cite{FZam4}, and then the series $\WA_n,\WB(0,n)$ and
$\WD_n$ \cite{FLuk2}. The operator product of such bosonic massless
free fields $X^i(z)$ takes the form

\eq
X^i(z) i\partial X^j(\z) = \frac{i\delta^{ij}}{z-\z} + \mno X^i(z)
i\partial X^j(\z) \mno
\;,
\en
where $\mno \mno$ denotes normal ordering. In ref. \cite{FLuk1},
Fateev and Lukyanov were
able to show that the $n$ fields $W^a(z)$ defined by
\eq
\sum_m W^m(z) (\a\partial)^{n+1-m} =
( i\partial X\cdd h^1 + \a\partial)
( i\partial X\cdd h^2 + \a\partial)
\ldots
( i\partial X\cdd h^{n+1} + \a\partial)
\;,\label{eq1}
\en
generate a closed operator product algebra, if $h^i$ are a basis of
$\Bbb R^n$ satisfying $h^i\cdd h^j = \delta^{ij} - 1/(n+1)$. This
generalises the well known construction for the Virasoro algebra in
terms of one free boson, and is known as the $\WA_n$ algebra. Such
closed expressions have not been found for any other series of
W--algebras, although partial results have been found for $\WD_n$ and
$\WB(0,n)$.
At this point we should make some remarks about notation. We take the
W--algebra $Wg$ to a `reductive' W--algebra (in the sense of
\cite{BWat1}) with basic fields whose conformal spin is one
greater than the exponents of $g$. Consequently we reserve the names
$WB_n, WC_n$ for W--algebras with purely bosonic fields of spin
$\{2,...,2n\}$. These are all well defined at the classical level, and
at the quantum level one must rely on the free field construction to
provide a proof of existence and consistency \cite{FFre5}.

\subsection{ Realisations of $\WA_n$ and $\WD_n$ in the Virasoro algebra
and free fields.}

These realisations were found by Lu et al. \cite{LPS},
based on the earlier work of Romans \cite{Roma1}. They used the
explicit form of the free field construction to deduce that one can
construct $Wg_n$ in terms of $Wg_{n-1}$ and one free field, for
$g=A,D$. We shall simply recap for the case $g=A$. Here, the
construction is given by \ref{eq1}. By considering the basis in which
$h^1$ is given by
\eq
h^1 = (0,0,..., -\sqrt{ n/n+1}) \;,
\en
the authors were able to show that one could express \wan in terms
of $\WA_{n-1}$ (constructed from the fields $X^1 \ldots X^{n-1}$ and
the field $X^n$. By recursively applying this argument they showed
that there is a realisation of \wan in terms of fields
$X^2\ldots X^n$ and the field $X^1$ where $X^1$ occurs only through
the Virasoro algebra generated by $ - (\partial X^1)^2  + \a\partial^2
X^1$. By applying a similar line of reasoning to
the field $f(z)$ which
is conjectured to generate all the fields in the $\WD_n$ algebra
\cite{FLuk2}, one
can show that there is a construction of $\WD_n$ in terms of a free
scalar field and $\WD_{n-1}$. Since $D_2 \equiv A_1 \oplus A_1$, the
final step in this recursion leads to two Virasoro algebras. Again,
since $D_3 \equiv A_3$, one can show that  for $n>2$ \wan can also be
realised in terms of $(n-2)$ free fields and two Virasoro algebras of
the same central charge.

They then applied these results to string theories based on
W--symmetry. We shall now reconsider these results in the light of our
new definition of a W--algebra.

\section{ Free field realisations and screening charges}

Our definition of the W--algebra free field realisation is in terms of
meromorphic conformal field theory \cite{Godd1}.
A meromorphic conformal field theory (mcft)
 consists of a Hilbert space $\cH$ and a vertex operator map from a
dense subspace $\cF$ of $\cH$ into the space of fields.
There is a distinguished state, the `conformal state'
$\vec L$, whose vertex operator is the stress-energy tensor of the theory, and
whose modes form a copy of the Virasoro algebra.

The important result for us is that
the vertex operator gives an isomorphism between states and local fields
via \hbox{$\vec\psi \to V(\psi,z)$}; from the axioms, it is possible
to deduce the uniqueness property that if $U_\phi(z)\vac =
e^{zL_{-1}}\vec\phi$, then $U_\phi(z)=V(\phi,z)$.
Since the singular part of the operator product of two fields only
includes fields of weight less than the sum of the two weights, to
show that there is a W--algebra of fields of spins $\Delta^i$ in a
mcft with a given Hilbert space, it is
only necessary to consider the space of states up to level
$2\max(\Delta) -1$.

It was quite necessary for the subsequent development of
the theory of the algebras that they commuted with certain
 screening charges, which are (formal) operators of the form
\eq
Q^j = \oint \frac{dz}{2\pi i}\mno \exp( i \beta \alpha^j \cdd
X(z))\mno
\;,
\en
where $\beta$ is a constant related to the central charge of the
W-algebra, and $\a^j$ runs over the simple roots of the associated
finite Lie algebra $g$.
However, it has proved more profitable to turn this around and instead
take this property as the definition of an mcft, which proves to be a
W--algebra \cite{FFre5}.
For each given
W--algebra one can show that the subspace of level
$2\max(\Delta)-1$ of the Hilbert space of the
W--algebra is isomorphic to the space of states annihilated by the
screening charges for all but a finite set of $c$-values.
Thus, we say that for all but a finite set of $c$--values,
the W--algebra corresponds to the set of fields which commute with the
screening charges, or by the field--state isomorphism, to the set of
states annihilated by the screening charges.
This is the analogue of Felder's BRST construction of the Virasoro
algebra \cite{Feld1}, and is discussed for $\WA_n$ by Mizoguchi and
Nakatsu in \cite{MNak1} and for general $g$ by Feigin and Frenkel in
\cite{FFre5}. This definition gives a closed W--algebra satisfying our
requirements concerning field content and is reductive, being a
quantum deformation of the classical case.
If, for all but a finite set of $c$--values, for level less than
$2\max(\Delta^i)$ the two Hilbert spaces $\cH, \cH'$ are isomorphic,
and the operator products of the two mcfts associated to them are
isomorphic to the same level, then we shall write
\eq
\cH \sim \cH'
\;.
\label{eq.sim}
\en
So, we shall take
 take the Hilbert space $\cH_{W}$ of the
W--algebra $Wg$ of a Lie algebra $g$ of rank $n$  as
\eq
\cH_W \sim (\cap_{\alpha\in\Sigma} ker_{Q^i}(\cH^{(n)} ) )
% / (\cup_{\alpha^i\Sigma} im_{Q^i}  )
\;.
\label{eq3}
\en
Whenever $c$ takes on of the finite set of values, one finds that
there are too many states in the kernel of the $Q^j$, corresponding to
the presence of extra null states in the W--algebra Verma module.

It is important to note
that the states in the free-field construction which are
annihilated by all the screening charges (modulo our restrictions on
level and $c$ above)  are
isomorphic to the states in an abstract W--algebra. Thus we can factor
any representation of a more complex W--algebra,
which uses only those states which are annihilated by a set of
screening charges, through a representation of the corresponding
W--algebra at that c--value.
The rest of this section will simply expand on this point and find
some consequences. We shall also use the field -- state isomorphism of
meromorphic conformal field theory and restrict our attention to the
states created by the W--algebra fields.

To proceed, we need some notation which we introduce now. For any
vector $\mu$, the field $\mu \cdd X / |\mu|$ satisfies the free field
operator product. We shall write the Hilbert space of this field as
$\cH_\mu$.
If $\mu^i$ are a basis of $\Bbb R^n$, then the Hilbert space of $n$
free bosons can be written as
\eq
\cH^{(n)} = \otimes \cH_{\mu^i}
\en
We shall  in particular be interested in the cases $\mu$ a simple root $\a^j$,
or fundamental weight $\lambda^j$ of a semi-simple Lie algebra.
These satisfy
\eq
\alpha^j \cdd \lambda^k = \delta^{jk} (\alpha^j)^2 /2
\;.\label{eq2}
\en
%Thus $\alpha^j \cdd \alpha^{\vee k} = C_{jk}$, the Cartan matrix of
%$g$.
We denote the set of simple roots of $g$ by $\Sigma$.
We shall also often denote the semi-simple Lie algebra whose simple roots
are given by $\Sigma$ by $g(\Sigma)$, and the corresponding W--algebra
by
$W(\Sigma)$.

As a first case, let us consider how a single free field in the
direction of a simple root $\alpha^{i_0}$ enters the expressions for
the W--algebra in the free field realisation.
Let us consider the basis of $\Bbb R^n$ given by
$%\eq
\{ \lambda^j, j\ne i_0 \} \cup \{ \alpha^{i_0} \}
\;.
$%\en
Then we can write the free field Fock space as
\eq
\cH^{(n)}
=
\cH_{\alpha^{i_0}} \otimes \cH^\perp
\;,
\;%\en
\hbox{ where }
\;%\eq
\cH^\perp
=
 \otimes_{j \ne i_0} \cH_{\lambda^j}
\;.
\en
We can now apply the results of Felder, \cite{Feld1}
which says that
for the states up to some level $2\Delta$, for all but a finite set of
$c$--values,
\eq
 ker_{Q^{i_0}} (\cH_{\alpha^{i_0}})
% / im_{Q^{i_0}}
\en
is isomorphic to the space of states $\cH_{L^{i_0}}$ generated by the
Virasoro algebra $L^{i_0}$ where
\eq
L^{i_o}(z) =
\mno
- 1 /( 2 |\a^{i_0}|^2 ) (\a^{i_0}\cdd\partial X(z))^2
 + (\beta - 2/(\beta |\a^{i_0}|^2)) \a^{i_0}\cdd\partial^2 X(z)
\mno
\;.
\en
(In fact if you look more closely at the construction we see that
these are a subset of those for which the W--algebra isomorphism
breaks down)
 From (\ref{eq2})  we also know that
\eq
Q^{i_0} \cH^\perp = 0
\;,
\label{eq.anh}
\en
And so combining (\ref{eq3}) and (\ref{eq.anh}) we obtain
\eq
\cH_W \sim
\left(
\cap_{\a^i\in\Sigma, i\ne i_0} ker_{Q^i}( \cH^{(n)} )
\right) \cap
\left(
\cH_{L^{i_0}} \otimes \cH^\perp
 \right)
 \;.
\en
Thus we see automatically that the free bosonic field $
(\a^{i_0}\cdd\partial X(z))/(|\a^{i_0}|) $
only occurs in $\cH_W$ via the Virasoro algebra $L^{i_0}$.
This provides our first result, that the free field in the direction
of {\em any } simple root appears only via its corresponding Virasoro
algebra $L^i$.
This generalises the results obtained by
\cite{LPS,Roma1} for the simple roots at the end of the
$A_n$ Dynkin diagrams, and those corresponding to the spinor and
spinor--bar representations of $D_n$ to all the simple root directions
of $g$.

If we regard this as a construction in the W--algebra of an isolated
simple root and $n-1$ free fields, we can see that
we can extend this result to the case of two (or more) mutually
orthogonal subsets of simple roots. We start by deleting one spot from
the Dynkin diagram of $g$ to leave two subsets of
simple roots, $\Sigma^1, \Sigma^2$. Let the deleted spot correspond to
the simple root $\a^{i_0}$,
and  we also denote the Fock space corresponding to the set of
root directions $\Sigma$ by $\cH^\Sigma$, so that
\eq
\cH^\Sigma = \otimes_{\alpha \in \Sigma}\cH_{\alpha}
\;.
\en
Then, given we can decompose the whole Fock space as
\eq
\cH^{(n)} =
\cH^{\Sigma^1} \otimes \cH_{\lambda^{ i_0}} \otimes
\cH^{\Sigma^2}
\;.
\en
Since
\eq
\alpha^j\cdd\alpha^k=0 \;\;,\; \alpha^j\in\Sigma^1,
\alpha^k\in\Sigma^2\;
\hbox{, and }\,
\alpha^j\cdd\lambda^{ i_0} =0 \;\;,\;
\alpha^k\in\Sigma^1, \alpha^j \in \Sigma^2
\;,
\en
we immediately see that
\eq
Q^i( \cH_{\lambda^{ i_0}} \otimes \cH^{\Sigma^1} ) = 0
\;\;,\;
\hbox{ for }\, \alpha^i \in\Sigma^2\;,
\en
and similarly replacing $1$ by $2$.
We then obtain
\eqq
\cH_{W(\Sigma)}
&\sim&
\ker_{Q^{i_0}}(\cH^{(n)})
\cap
\left(
\cap_{\alpha^i \in \Sigma^1} ker_{Q^i}( \cH^{(n)} )
\right)
\cap
\left(
\cap_{\alpha^i \in \Sigma^2} ker_{Q^i}( \cH^{(n)} )
\right)
\nonumber\\
&=&
\ker_{Q^{i_0}}(\cH^{(n)})
\cap
\left[
\left(
\cap_{\alpha^i \in \Sigma^1} ker_{Q^i}( \cH^{\Sigma^1})  \right)
\otimes \cH_{\lambda^{i_0}} \otimes \cH^{\Sigma^2} \right]
\nonumber\\
&&\qquad
\qquad
\cap
\left[
 \cH_{\lambda^{i_0}} \otimes \cH^{\Sigma^1} \otimes
\left(
\cap_{\alpha^i \in \Sigma^2} ker_{Q^i}( \cH^{\Sigma^2})  \right)
\right]
\nonumber\\
&\sim&
\ker_{Q^{i_0}}(\cH^{(n)})
\cap
\left[
\cH_{W(\Sigma^1)} \otimes \cH_{\lambda^{i_0}} \otimes \cH_{W(\Sigma^2)}
\right]
\enn
Again, if we are careful about the $c$--values for the various
W--algebras, and use the determinant formulae for the W--algebras, we
can see that the $c$--values for which there are null states for the
smaller W--algebras at levels less than $2\Delta$ are a subset of
those of the larger.

The final upshot is that the free fields in the directions spanned by
the simple roots in the sets $\Sigma^1, \Sigma^2$ only appear in the
final realisation of $W(\Sigma)$ through the W--algebras $W(\Sigma^1)$
and $W(\Sigma^2)$ respectively.
Since the spaces $\Sigma_{W(\Sigma^i)}$ are isomorphic to the vacuum
representations of the W--algebras $W(\Sigma^i)$, this means that we
can replace them by the abstract W--algebra, subject to the condition
that these have central charges $c^i$ given by
\eq
c^i = n^i - (\beta\rho^i - 1/\b \rho^{\vee i})^2
\;,
\label{eq.cvals}
\en
where $n^i = \rank(g(\Sigma^i))$, $\rho^i = \sum_{\a^i\in\Sigma^i}
\lambda^i$,
$\rho^{\vee i} =  \sum_{\a^i\in\Sigma^i} \lambda^{\vee i}$.

This now provides us with our second result, which is that one can
find a realisation of a W--algebra $W(\Sigma)$ in the W--algebras
$W(\Sigma^i)$ for subdiagrams obtained by deleting one node, and one
free field, subject to the condition that the c-values of the three
W--algebras all satisfy (\ref{eq.cvals}) for some $\beta$.

\subsection{Example}

We run through the simple example of $\WA_4$ in the free field
construction and show that we can find a realisation  in terms of
$\WA_2$, $\WA_1$ and one free field.
%LONG EXAMPLE OF $\WA_4$ to go in here

The Dynkin diagram of $\WA_4$ is simply
\vskip -.2cm
\begin{center}
\begin{picture}(30,20)(-5,-5)
\dynkina{4}
\end{picture}
\end{center}

If we number the simple roots from $1$ to $4$ from left to right, then
we shall consider deleting the root direction $3$, and take
a basis of $\Bbb R^4$
\eq
\{ \alpha^1, \a^2, \lambda^3, \a^4 \}
\;.
\en
If we consider the vectors $h^i$ arising in the free field
construction (\ref{eq1}), these can be written in terms of the basis
vectors as
\eq
3h^1 = \l^3 + 2\a^1 + \a^2
\;,\;
3h^2 = \l^3 - \a^1 + \a^2
\;,\;
3h^3 = \l^3 - 2\a^2 - \a^1
\;,\;
\nonumber
\en\eq
2h^4 = \a^4 - \l^3
\;,\;
2h^5 = -\a^4 -\l^3
\;.
\en
If we insert these into (\ref{eq1}), we obtain
\eq
\sum_m W^m(z) (\a\partial)^{n+1-m} =
\nonumber
\en
\eq
\Bigl(( W_{1,2} + \mu L_{1,2} + \mu^3
+ 3\a \mu\mu' + \a^2\mu'' )
 +
( L_{1,2} + 3 \mu^2 + 3\a\mu')\a\partial
+ (3\mu) (\a\partial)^2
+ (\a\partial)^3
\Bigr)
\en
\nonumber
\eq
\times
\Bigl(
( (9/2)\mu^2 - (3\a/2)\mu' - L_4 )
-3\mu\a\partial + (\a\partial)^2
\Bigr)
\;,
\en
where
\eqq
W_{1,2} + L_{1,2}(\a\partial) + (\a\partial)^3
&=&
( i\partial X\cdd(2\a^1 + \a^2)/3 + \a\partial )
( i\partial X\cdd(-\a^1 + \a^2)/3 + \a\partial )
\cont\times
( i\partial X\cdd(-\a^1 -2\a^2)/3 + \a\partial )
\;,
\nonumber \\
L_{4} &=&  - (\a^4\cdd\partial X / |\a^4| )^2
+ i\a\a^4\cdd\partial^2 X
\;,\\
\mu &=& (1/3)\lambda^3\cdd i\partial X \;.
\nonumber
\enn
As can be seen, this provides a construction in $\WA_2, \WA_1$ and one
free field, with the central charges satisfying (\ref{eq.cvals}),
where $\a = \b - 1/\b$.

\section{Conclusions}

By repeating this procedure outlined above, we can construct a
realisation of a W--algebra $W(\Sigma)$ in the W--algebras
$W(\Sigma^i)$ for subdiagrams obtained by deleting nodes from the
Dynkin diagram of the {\em
finite} algebra, and one
free field for each node deleted, subject to the condition that the
c-values of the W--algebras all satisfy (\ref{eq.cvals}) for some
$\beta$.

This is very reminiscent of the classification of regular subgroups of
a semisimple Lie group. In that case there is a regular subgroup for
of the form
\eq
 G(\Sigma^1) \otimes ... \otimes G(\Sigma^n) \otimes U(1)^{(n-1)}
\;,
\en
where $\{\Sigma^i\}$ are the simple roots of sub-diagrams obtained by
deleting $n$ spots from the {\em affine}
Dynkin diagram for $g$.
Let us now take the point of view that
W--algebra is the symmetry algebra obtained by Quantum Hamiltonian
reduction of a  certain coadjoint orbit of the affine Lie group,
by a subgroup corresponding to the nilpotent subalgebra.
In refs. \cite{Prin}, de Groot et al. considered a similar
system in which they
associated an integrable hierarchy to each affine Lie algebra, choice
of grading of the affine root space, and choice of regular element of
a particular subspace, by a Hamiltonian reduction with respect to a
nilpotent subalgebra of the affine algebra. They then found that they
could find different spaces which all had the same integrals of motion
by a set of partial gauge fixings of the reduction, which were given
by gradings of the root space which were between the grading they
first thought of and another in a certain partial ordering.
The case we have here of a W--algebra can be certainly be viewed as a
Hamiltonian reduction, and the grading to which it is associated is
the grading inherited from the principal $sl(2)$ embedding. This
assigns grade 1 to each simple root. There is a partial gauge fixing
for each grade which assign either 1 or 0 to each simple root. The
occurence of the realisations we have found can be viewed as a
consequence of this property; we can partially gauge fix and so reduce
our effective space to the space of the sub-W--algebra and the free
fields before completing the process to arrive at the W--algebra.

One can also turn to the affine equivalent of this construction.
The commutant of the screening charges associated to all the simple
roots in an affine diagram now give the local conserved quantities of
motion of the Affine Toda Field Theory \cite{FFre4}. The analysis
presented here
would say that one can find a construction of these local conserved
quantities in terms of the W--algebras for any regular subalgebra of
$g$ and the appropriate number of free fields.

If one wishes to go on to consider the more general W--algebras
obtained recently \cite{Int}, then one needs to have a better
understanding of the quantum screening charges for these cases, which
have the classical form
$Q^\beta =  \sum_{\alpha\in\Sigma} \int
\Tr (G(x) E_\alpha G(x)^{-1} E_{-\beta} ) $
where $G$ is a WZW field, and $\Sigma$ is a set of simple roots for
the nilpotent subalgebra being gauged.

\section{Acknowledgements}

I would like to thank P. Bowcock for several illuminating
conversations, and H.G. Kausch for discussions over a long period.
This work was supported by an SERC research assistantship

%the end.

%\bibliography{bib1}

\begin{thebibliography}{10}

\bibitem{Zamo1}
A.~B. Zamolodchikov,
\newblock Theoretical and Mathematical Physics 65 (1985) 347.

\bibitem{FLuk1}
V.~A. Fateev and S.~L. Luk'yanov,
\newblock Int. J. Mod. Phys. A3 (1988) 507.

\bibitem{FLuk2}
V.~A. Fateev and S.~L. Luk'yanov,
\newblock Sov. Sci. Rev. A15 (1990) 1.

\bibitem{BBSS}
F.~A.~Bais, P.~Bouwknegt, K.~Schoutens and M.~Surridge, Nucl. Phys. B304 (1988)
  348; Nucl. Phys. B304 (1988) 371.

\bibitem{BFFOW}
J.~Balog, L.~Feh\'er, P.~Forg\'acs, L.~{O' Raifeartaigh} and A.~Wipf, \newblock
  Annals of Physics 203 No. 1 (1990) 76. \newblock Phys. Lett. B244 (1990) 435.

\bibitem{Roma1}
L.~J. Romans,
\newblock Nucl. Phys. B352 (1991) 829.

\bibitem{LPS}
H.~Lu, C.~N. Pope, S.~Schrans and K.~W. Xu{\it, \newblock The Complete Spectrum
  of the $W_N$ String}, \newblock Texas A and M Preprint CTP TAMU--5/92 (1992).
  H.~Lu, C.~N. Pope, S.~Schrans and X.~J. Wang{\it, \newblock New Realisations
  of W algebras and W strings}, \newblock Texas A and M Preprint CTP
  TAMU--15/92 (1992); {\it, \newblock Sibling and Exceptional $W$ Strings},
  \newblock Texas A and M Preprint CTP TAMU--10/92 (1992).

\bibitem{Watt3}
G.~M.~T. Watts,
\newblock Phys. Lett. B245 (1990) 65.

\bibitem{Bouw3}
P.~Bouwknegt,
\newblock Phys. Lett. 207B (1988) 295.

\bibitem{BFKNRV1}
R.~Blumenhagen, M.~Flohr, A.~Kliem, W.~Nahm, A.~Recknagel and R.~Varnhagen{\it,
\newblock W-algebras with two and three Generators},
\newblock Nucl. Phys. B361 (1991) 255 .

\bibitem{KWat1}
H.~G. Kausch and G.~M.~T. Watts,
\newblock Nucl. Phys.  (1991) 740.

\bibitem{FZam4}
V.~Fateev and A.~Zamolodchikov,
\newblock Nucl. Phys. B280 [FS18] (1987) 644.

\bibitem{BWat1}
P.~Bowcock and G.~M.~T. Watts{\it,
\newblock On the Classification of Quantum W--algebras},
\newblock Enrico Fermi Institute Preprint EFI--91--63 (1991),
\newblock To appear in Nucl. Phys. B.

\bibitem{FFre5}
B.~Feigin and E.~Frenkel{\it,
\newblock Affine Kac--Moody algebras at the critical level and Gelfand--Dikii
  algebras},
\newblock Research Institute in Mathematical Sciences, Kyoto, Preprint RIMS 796
  (1991).

\bibitem{Godd1}
P.~Goddard,
\newblock Meromorphic Conformal Field Theory,
\newblock {\em in\/:} Infinite Dimensional Lie Algebras and Lie Groups,  ed.
  V.~G. Kac, World Scientific, 1989,
\newblock CIRM-Luminy July conference on Infinite dimensional Lie Algebras and
  Lie Groups, Marseille 1988.

\bibitem{Feld1}
G.~Felder{\it,
\newblock BRST Approach to Minimal Models},
\newblock Nucl. Phys. B317 (1989) 215.

\bibitem{MNak1}
S.~Mizoguchi and T.~Nakatsu{\it,
\newblock BRST Structure of the $W_3$ minimal model},
\newblock University of Tokyo Preprint UT-566 (1991).

\bibitem{Prin}
M.~F.~de~Groot, T.~J.~Hollowood and J.~L.~Miramontes, {\it Generalised Drinfeld
  Sokolov Hierarchies I,} IASSNS-HEP-91/19; N.~Burroughs, M.~F.~de~Groot,
  T.~J.~Hollowood and J.~L.~Miramontes, {\it Generalised Drinfeld Sokolov
  Hierarchies II,} IASSNS-HEP-91/42; N.~Burroughs, {\it Coadjoint orbits of the
  generalised sl(2), sl(3) KdV hierarchies}, IASSNS-HEP-91/67.

\bibitem{FFre4}
B.~L. Feigin and E.~V. Frenkel{\it,
\newblock Free Field resolutions and affine Toda theory},
\newblock Research Institute in Mathematical Sciences, Kyoto, Preprint
  RIMS--827 (1991).

\bibitem{Int}
L.~Feh\'er, L.~O'Raifeartaigh, P.~Ruelle, I.~Tsutsui and A.~Wipf, {\it
  Generalized Toda theories and W algebras associated with integral gradings},
  Ann. Phys. 213 (1992) 1--20; L.~Feh\'er, {\it W-Algebras of generalized Toda
  theories}, Dublin preprint {DIAS-STP-91-22}; F.~A.~Bais, T.~Tjin and
  P.~van~Driel, {\it Covariantly coupled chiral algebras}, Nucl. Phys. B357
  (1991) 632.

\end{thebibliography}

\end{document}